%Paper: hep-th/9511060
%From: dowker@a3.ph.man.ac.uk
%Date: Thu, 09 Nov 1995 14:28:29 GMT
%Date (revised): Thu, 23 Nov 1995 09:54:51 GMT

\input jytex.tex   % available from hep-th
%\draft
\typesize=10pt
\magnification=1200
\baselineskip=17truept
\hsize=6truein\vsize=8.5truein
%\leftmargin=1.25in
%\oddleftmargin=.5in
%\evenleftmargin=1.5in
\sectionnumstyle{blank}
\chapternumstyle{blank}
\chapternum=1
\sectionnum=1
\pagenum=0
% title style follows

\def\begintitle{\pagenumstyle{blank}\parindent=0pt\begin{narrow}[0.4in]}
\def\endtitle{\end{narrow}\newpage\pagenumstyle{arabic}}

% exercise style follows

\def\beginexercise{\vskip 20truept\parindent=0pt\begin{narrow}[10
truept]}
\def\endexercise{\vskip 10truept\end{narrow}}

% **************    my jyTeX abbreviations   *****************

\def\eql#1{\eqno\eqnlabel{#1}}
\def\ref{\reference}
\def\peq{\puteqn}
\def\pref{\putref}

\def\mgn{\marginnote}
\def\bex{\begin{exercise}}
\def\eex{\end{exercise}}
\def\dst{\displaystyle}
% *********************** My definitions ************************

\font\open=msbm10 %scaled\magstep1 % For VAX. Borde p195.
 %scaled\magstep1 % For VAX. Borde p195.
%\font\open=msym10 %scaled\magstep1 % For Arbortxt on PC
%\font\opens=msym8 %scaled\magstep1 % For Arbortxt on PC
%\font\goth=eufm10  % For Arbortxt on PC, and VAX. Borde p199
%\font\ssb=cmss10
%\font\smsb=cmss8
\def\mbox#1{{\leavevmode\hbox{#1}}}
\def\hspace#1{{\phantom{\mbox#1}}}
\def\oR{\mbox{\open\char82}}

\def\rS{{\rm S}}

 %in jyTeX
 %in jyTeX
\def\bga{{\bmit\gamma}} %in jyTeX
 %in jyTeX
 %in jyTeX
% in jyTeX

\def\ga{\gamma}
\def\de{\delta}
\def\Ga{\Gamma}

\def\la{\lambda}

\def\Om{\Omega}

\def\ze{\zeta}

\def\zf{$\zeta$--function}
\def\zfs{$\zeta$--functions}

     % Newline

\def\frac#1/#2{\leavevmode\kern.1em
\raise.5ex\hbox{\the\scriptfont0 #1}\kern-.1em/\kern-.15em
\lower.25ex\hbox{\the\scriptfont0 #2}}
\def\sfrac#1/#2{\leavevmode\kern.1em
\raise.5ex\hbox{\the\scriptscriptfont0 #1}\kern-.1em/\kern-.15em
\lower.25ex\hbox{\the\scriptscriptfont0 #2}}

\def\gtorder{\mathrel{\raise.3ex\hbox{$>$}\mkern-14mu
             \lower0.6ex\hbox{$\sim$}}}
\def\ltorder{\mathrel{\raise.3ex\hbox{$<$}\mkern-14mu
             \lower0.6ex\hbox{$\sim$}}}

\def\semidirprod{\rlap{\ss C}\raise1pt\hbox{$\mkern.75mu\times$}}
\def\for{\lower6pt\hbox{$\Big|$}}
\def\fish{\kern-.25em{\phantom{abcde}\over \phantom{abcde}}\kern-.25em}

 %triple
%dot
 %double
%dot
 %double dot
%for small #1

\def\boxit#1{\vbox{\hrule\hbox{\vrule\kern3pt
        \vbox{\kern3pt#1\kern3pt}\kern3pt\vrule}\hrule}}
\def\dalemb#1#2{{\vbox{\hrule height .#2pt
        \hbox{\vrule width.#2pt height#1pt \kern#1pt
                \vrule width.#2pt}
        \hrule height.#2pt}}}

        %double stroke
\def\frac#1#2{{{#1}\over{#2}}}
 %lower covariant deriv.
    %lower ordinary  deriv.

\def\noin{\noindent}

      %Connection
    %Connection'
\def\comb#1#2{{\left(#1\atop#2\right)}}

\def\eg{{\it e.g. }}

 %gives average <#1>
 %gives thermal average <<#1>>
   %gives bracket <#1|#2>
 %gives big bracket <#1|#2>
  %gives
%matrix element <#1|#2|#3>

  %

\def\3j#1#2#3#4#5#6{\left\lgroup\matrix{#1&#2&#3\cr#4&#5&#6\cr}
\right\rgroup}

\def\man{{\cal M}}

\def\bga{{\bmit\ga}}
\def\m?{\mgn{?}}

%  *******************  Journal refs **********************

\def\aop#1#2#3{{\it Ann. Phys.} {\bf {#1}} (19{#2}) #3}

\def\cqg#1#2#3{{\it Class. Quant. Grav.} {\bf {#1}} (19{#2}) #3}

\def\jmp#1#2#3{{\it J. Math. Phys.} {\bf {#1}} (19{#2}) #3}
\def\jpa#1#2#3{{\it J. Phys.} {\bf A{#1}} (19{#2}) #3}

\def\np#1#2#3{{\it Nucl. Phys.} {\bf B{#1}} (19{#2}) #3}
\def\pl#1#2#3{{\it Phys. Lett.} {\bf {#1}} (19{#2}) #3}

\def\prD#1#2#3{{\it Phys. Rev.} {\bf D{#1}} (19{#2}) #3}

\def\cras#1#2#3{{\it Comptes Rend. Acad. Sci. (Paris)} {\bf{#1}} (#2) #3}

\def\mpcps#1#2#3{{\it Math. Proc. Camb. Phil. Soc.} {\bf{#1}} (19{#2}) #3}

\def\aim#1#2#3{{\it Adv. in Math.} {\bf {#1}} (19{#2}) #3}

\def\invm#1#2#3{{\it Invent. Math.} {\bf {#1}} (19{#2}) #3}

% *******************   Main text *********************
\begin{title}
\vglue 20truept
\righttext {MUTP/95/21}
\righttext{hep-th/9511060}
\leftline{\today}
\vskip 30truept
\centertext {\Bigfonts \bf Spectral invariants for the Dirac}
\vskip 5truept
\centertext {\Bigfonts \bf  equation on the d-ball with}
\vskip 5truept
\centertext {\Bigfonts \bf various boundary conditions}
%\vskip 5truept
%\centertext {\Bigfonts \bf}
\vskip 20truept
\centertext{J.S.Dowker\footnote{dowker@a3.ph.man.ac.uk},
 J.S.Apps\footnote{jsa@a3.ph.man.ac.uk}}
\vskip 7truept
\centertext{\it Department of Theoretical Physics,\\
The University of Manchester, Manchester, England}
\vskip10truept
%\centertext{and}
\vskip10truept
\centertext {K.Kirsten\footnote{kirsten@tph100.physik.uni-leipzig.de},
 M.Bordag\footnote{bordag@qft.physik.uni-leipzig.d400.de}}
\vskip7truept
\centertext{\it Universit\"at Leipzig, Institut f\"ur Theoretische
Physik,\\ Augustusplatz 10, 04109 Leipzig, Germany}
\vskip 20truept
\centertext {Abstract}
\begin{narrow}
The mode properties for spectral and mixed boundary conditions
for massless spin-half fields are derived for the $d$--ball. The corresponding
functional determinants and traced heat-kernel coefficients are presented,
the latter as polynomials in $d$.
\end{narrow}
\vskip 5truept
\righttext {November 1995}
\vskip 60truept
\righttext{Typeset in \jyTeX}
\vfil
\end{title}
\pagenum=0
\section{\bf 1. Introduction.}

In putative supersymmetric field theories on manifolds with boundary, the
question arises of the boundary conditions satisfied by the higher-spin
fields.
These problems are encountered for example in quantum cosmology
[\pref{DandH, DandE, DandE2, Esposito, Barv, MandP, Moss, Kam, KandM}] and
have become more pressing recently, particularly in gauge theories
[\pref{Kam2, MandV, Esposito2}].

It is generally assumed that the conditions
should be such as to make the relevant operators
self-adjoint (see \eg [\pref{ETP}]). One possibility is the spectral
condition introduced by Atiyah, Patodi and Singer [\pref{APS}] in their
extension of the spin-index theorem to the non-empty boundary case.
Although it seems that spectral conditions are not suitable for
supersymmetry, they are of undoubted interest beyond this particular purpose.

In this paper we report on the calculation of important quantities in
the spectral geometry of spin-1/2 fields on the $d$-ball with these nonlocal
boundary conditions, namely the integrated heat-kernel asymptotic
expansion coefficients and the functional determinants. For comparison,
we also treat the case of mixed (local) conditions, which are possibly
of more supersymmetric significance.

Although the $d$-ball is a very particular manifold, it turns out in the
corresponding scalar Dirichlet and Neumann cases that the results are
surprisingly restrictive of the {\it general} form of the heat-kernel
expansion [\pref{BGV}]. One of the motivations for the present calculation is
to prepare the way for a similar discussion with spinors.
\section{\bf2. Spinor modes on the d-ball. Spectral conditions.}

The eigenvalue Dirac equation on the Euclidean d-ball is
$$
-i\Ga^\mu\nabla_\mu\psi_{\pm}=\pm k\psi_{\pm},
\quad\Ga^{(\mu}\Ga^{\nu)}=g^{\mu\nu},
\eql{Dirac}$$
and the nonzero modes are separated in polar coordinates,
$ds^2=dr^2+r^2d\Om^2$, in standard fashion to be regular at the origin,
($A$ is a radial normalisation factor),
$$\eqalign{
&\psi_{\pm}^{(+)}={A\over r^{(d-2)/2}}\comb{iJ_{n+d/2}(kr)\,Z^{(n)}_+(\Om)}
{\pm J_{n+d/2-1}(kr)\,Z^{(n)}_+(\Om)}\cr
\noalign{\vskip0truept}
&\psi_{\pm}^{(-)}={A\over r^{(d-2)/2}}\comb{\pm J_{n+d/2-1}(kr)\,Z^{(n)}_
-(\Om)}{iJ_{n+d/2}(kr)\,Z^{(n)}_-(\Om)}.\cr}
\eql{diracmodes} $$
Here the $Z^{(n)}_{\pm}(\Om)$ are the well-known spinor modes on the \mgn{REFS}
unit $(d-1)$--sphere (some modern references are [\pref{AandT, JandK, CandH}])
satisfying the intrinsic equation
$$
-i\ga^j\widetilde\nabla_j Z^{(n)}_{\pm}=\pm\la_n Z^{(n)}_{\pm},
\eql{spheig}$$where
$$
\la_n=\big(n+{d-1\over2}\big),\quad n=0,1,\ldots\,\,.
$$
Each eigenvalue is greater than $1/2$ and has degeneracy
$$
{1\over2}d_s\,\comb{d+n-2}n.
$$
The dimension, $d_s$, of $\psi$--spinor space is $2^{d/2}$ if $d$ is
even. For odd $d$ it is $2^{(d+1)/2}$ and has
been doubled in order to implement the boundary conditions. Appendix A
contains a more systematic discussion of $\ga$-matrices and spinors.

The projected $\ga$-matrices are given by
$$
\Ga^r=\left(\matrix{{\bf0}&{\bf 1}\cr{\bf1}&{\bf0}\cr}\right),\quad
\Ga^j=\left(\matrix{{\bf0}&i\ga^j\cr-i\ga^j&{\bf0}\cr}\right),\quad\Ga^5=
\left(\matrix{{\bf1}&{\bf 0}\cr{\bf0}&-{\bf1}\cr}\right).
\eql{matrices}$$

Spectral boundary conditions are applied, effectively as in D'Eath
and Esposito [\pref{DandE2}], by setting the negative (positive) $Z$-modes
of the positive (negative) chirality parts of $\psi$, to zero at $r=1$,
the other modes remaining free. This leads to the condition
$J_{n+d/2-1}(k)=0$.

Roughly speaking, spectral conditions amount\mgn{Redbook D 7,8} to
requiring that zero-modes of (\peq{Dirac}) should be
square-integrable on the elongated manifold obtained from the ball by
extending the narrow collar (of approximate, product metric $dr^2+d\Om^2$)
just inside the surface to values of $r$ ranging from $1$ to $\infty$. This
will be so if the modes of $A=\Ga^r\Ga^j\nabla_j$ with {\it negative}
eigenvalues are suppressed at the boundary,
(\eg [\pref{APS, Gilkey2, FORW, NandT, RandS, Ma, MandS,NandS,Gilkey3}] ).
At $r=0$ the modes vanish except that with $n=0$ which has the opposite
handedness.

{}From (\peq{spheig}) and (\peq{matrices}), the boundary operator, $A_0$,
is $A_0=\Ga^r\Ga^j\nabla_j\big|_{r=1}=$ \break
$\big(\Ga^5\otimes-i\ga^j\big)\big({\bf1}\otimes\widetilde\nabla_j\big)=
\Ga^5\otimes\big(-i\ga^j\widetilde\nabla_j\big)$ and its eigenstates are
$$
A_0\comb{Z^{(n)}_+}{Z^{(n)}_-}=\la_n\comb{Z^{(n)}_+}{Z^{(n)}_-},\quad
A_0\comb{Z^{(n)}_-}{Z^{(n)}_+}=-\la_n\comb{Z^{(n)}_-}{Z^{(n)}_+}.\quad
\eql{bop}$$Then,
from (\peq{diracmodes}), we see that the negative modes of $A_0$ are
associated with the radial factor $J_{n+d/2-1}(kr)$, hence the condition
quoted above.

We put $p=n+d/2-1$ making the implicit eigenvalue equation,
$$
J_p(k)=0
\eql{imp}$$
with degeneracies
$$
N_p^{(d)}={d_s\over(d-2)!}\big(p-{d\over2}+2\big)
\big(p-{d\over2}+3\big)\ldots\big(p+{d\over2}-1\big)
\eql{spdegen}$$
where $p\ge d/2-1$ and is integral for even $d$ but half odd-integral for
odd $d$. The form of the degeneracies shows that $p$ can start at 1 if
$d$ is even and at $1/2$ if $d$ is odd. For $d=4$, we obtain agreement with
D'Eath and Esposito [\pref{DandE2}]. The normalisation in
(\peq{diracmodes}) is $A=\big(J_{n+d/2}(k)\big)^{-1}$.

The case of the disc, $d=2$, needs special treatment.
The implicit equation is still (\peq{imp}),
with $p=1,2,\ldots$, but the degeneracy is just 2.
\mgn{Check this}
\section{\bf 3. Mixed boundary conditions.}

For mixed boundary conditions, [{\pref{MandP, Gilkey3, BandG2, MandL, Luck,
Mcav}], we apply $P_+\psi=0$ at $r=1$ where the projection is
$$
P_+={1\over2}\big({\bf 1}-i\Ga^5\Ga^\mu\,n_\mu\big)
\eql{mproj}$$in terms of the inward normal $n_\mu$.

For the geometry of the ball
$$
P_+={1\over2}\left(\matrix{{\bf1}&i{\bf1}\cr-i{\bf1}&{\bf1}\cr}\right)
$$and so for $\psi^{(+)}_\pm$,
$$
J_{n+d/2}(k)=\mp J_{n+d/2-1}(k)
$$
and for $\psi^{(-)}_\pm$,
$$
J_{n+d/2-1}(k)=\mp J_{n+d/2}(k),\quad n=0,1,2,\ldots.
$$

Thus, taking $p=n+d/2$, the implicit eigenvalue equation is as in
[\pref{DandE}],
$$
J_p^2(k)-J_{p-1}^2(k)=0
\eql{mimp}$$
while the degeneracies are
$$
N_p^{(d)}={d_s\over2(d-2)!}\big(p-{d\over2}+1\big)
\big(p-{d\over2}+2\big)\ldots\big(p+{d\over2}-2\big)
\eql{mdegen}$$
where $p\ge d/2$ and is integral for even $d$ but half odd-integral for
odd $d$. The form of the degeneracies shows again that $p$ can start at 1
if $d$ is even and at $1/2$ if $d$ is odd. In two dimensions, the
degeneracy is unity.

For both conditions, the \zf\ is $\ze_d(s)=\sum_p\sum_{k_p}
N_p^{(d)}(k_p)^{-2s}$, $k_p$ being the positive roots of (\peq{imp})
or of (\peq{mimp}). The functional determinant is
$\exp\big(-\ze_d'(0)\big)$ and the traced heat-kernel expansion, $K(\tau)=
\sum_{n=0,1/2,\ldots} B_n\,\tau^{n-d/2}$.
\section{\bf 4. Polynomial form of traced heat-kernel coefficients.}

Specific, integral forms exist for the first few {\it local}
coefficients, [\pref{MandL, BandG2, BGV}], which, computed on the
$d$-ball, give
$$\eqalign{
%&B_0^{(L)}(d)={d_s\over2^{d-1}\Ga(d/2)},\cr
&B_0^{(L)}(d)={2^{-d-1}d_s\over\Ga(1+d/2)},\cr
&B_{1/2}^{(L)}(d)=0,\cr
&B_1^{(L)}(d)=-{2^{-d}d_s\over6\,\Ga(d/2)}\,(d-1),\cr
&B_{3/2}^{(L)}(d)={2^{-d}d_s\sqrt\pi\over64\,\Ga(d/2)}\,(d-1)\,(d-3),\cr
&B_2^{(L)}(d)={2^{-d}\,d_s\over3780\,\Ga(d/2)}\,(d-1)\,(d+3)\,(17d-46).\cr}
\eql{lintforms}$$

For particular $d$'s, (\peq{lintforms}) is consistent with the results
obtained from (\peq{mimp}) and (\peq{mdegen}) using the method described
in Bordag {\it et al} [\pref{BEK}]. The individual values following from this
calculation are not displayed here since they are better used to construct
the polynomial content of the coefficients, some higher examples of which are
exhibited in Appendix B. There is no difficulty in finding any coefficient.

Turning to the spectral case, although there appears to be no
known general forms corresponding to those for local coefficients,
polynomial expressions can be obtained in the present geometry.
These are written conjecturally as
$$\eqalign{
B_n^{(S)}(d)&=2^{-d}d_s\bigg({\overline F_n(d)\over\Ga\big((d+1)/2\big)}
+\sqrt\pi\,{\overline G_n(d)\over\Ga\big(d/2\big)}\bigg),\quad n=1/2,3/2,\ldots
\cr
&=2^{-d}d_s\bigg({\overline F_n(d)\over\Ga\big(d/2\big)}+\sqrt\pi\,{
\overline G_n(d)\over\Ga\big((d+1)/2\big)}\bigg),\quad n=0,1,\ldots\cr}
\eql{lform}$$
where $\overline F_n$ and $\overline G_n$ are polynomials of degree
$2n-1$. For $n\ge1$ a factor of $d-1$ is extracted,
$\overline F_n=(d-1)F_n$, $\,\overline G_n=(d-1)G_n$ and the $F$
and $G$ fitted using specifically evaluated coefficients over several
dimensions. This yields
$$\eqalign{
&\overline F_0(d)={1\over d},\quad\overline G_0(d)=0,\cr
&\overline F_{1/2}(d)={1\over2},\quad \overline G_{1/2}(d)=-{1\over2},\cr
&F_1(d)={1\over3}\,,\quad G_1(d)=-{1\over4}\,,\cr
&F_{3/2}(d)={1\over24}\,(4d-11)\,,\quad
G_{3/2}(d)=-{1\over192}\,(7d-17)\,,\cr
&F_2(d)={1\over945}\,(d-6)\,(5d-13)\,,\quad
G_2(d)=-{1\over384}\,(d-6)\,(7d-20)\,.\cr}
\eql{spoly}$$
Further polynomials are given in Appendix C. The forms have been checked
to $d=19$. The coefficients for $d=4$ were also given earlier by
Kirsten and Cognola [\pref{KandC}].

We remark on the circumstance that alternate spectral coefficients
(depending on the dimension) are comprised
of two parts, one proportional to $\sqrt\pi$ and the other to $1/\sqrt
\pi$. By contrast, for local (mixed) boundary conditions there are no
$1/\sqrt\pi$ terms and this would be the expected behaviour.

A similar structure to (\peq{lform}) is encountered in the case of
\mgn{Why? Elliptic? Local?} local conditions for physical components of
{\it higher} spin fields in four dimensions, [\pref{KandC}].

When, as here, the manifold is not product near the boundary, the
spectral asymptotic expansion has been established by Grubb [\pref{Grubb}]
and by Grubb and Seeley [\pref{GandS1}]. In the product, cylindrical case
Grubb and Seeley give a construction of the \zf\ in terms of the \zfs\
on the doubled manifold and on the boundary which would
yield a structure for the heat-kernel somewhat akin to (\peq{lform}).

It should also be remarked that, in the general case, if $d$ is even
there can be logarithmic terms in the heat-kernel expansion, equivalent to
double poles in the \zf. These are absent here, the reason possibly
being that the heat-kernel expansion for a massless Dirac field on the
odd dimensional boundary, $\rS^{d-1}$, terminates with the $\tau^{-1/2}$ term,
a well known fact. This mechanism is explicit for the even $d$-hemisphere
using Grubb and Seeley's product construction.

The values (\peq{spoly}) show, in particular, that the massless spin-1/2
scaling behaviour is governed in the spectral case by the numbers,
$$
\matrix{\ze_2(0)=-\dst{1\over12},&\ze_3(0)=0,&\ze_4(0)
=\dst{11\over360},&\ze_5(0)=0,\cr
\noalign{\vskip10truept}
\ze_6(0)=-\dst{191\over15120},&\ze_7(0)=0,&\quad\ze_8(0)
=\dst{2497\over453600},&etc.\cr}
\eql{zezero}$$
\vskip5truept
\noin which equal those for local (or mixed) boundary conditions as
was noted by D'Eath and Esposito [\pref{DandE2}] in four dimensions. The mixed
values also follow from those on the $d$-hemisphere by conformal invariance
which may account for the equality since the Grubb-Seeley formula shows that
on the hemisphere $\ze_d^{(S)}(0)=\ze_d^{(L)}(0)$, each being half the full
sphere value.

In addition we note the result,
$$
B_{d/2-1}^{(S)}(d)=0,\quad d\,\,{\rm even}
\eql{van}$$
which corresponds to the vanishing residue of the pole of the
spectral \zf\ at $s=1$.

%In the present
%case the spectrum, (\peq{bop}), of the boundary Dirac operator $A$ is
%symmetrical about the origin and so the $\eta$-function contribution to
%$\ze(s)$ vanishes.
\newpage
\noindent{\bf5. Spectral functional determinants.}

Application of the techniques fully described in our earlier works
\mgn{SPECTRAL.MTH} [\pref{BGKE,Dow8,Dow10}] leads straightforwardly to
$$\eqalign{
&\ze_2'(0)=2\ze_R'(-1)+{2\over3}\ln2+{5\over12},\cr
\noalign{\vskip5truept}
&\ze_3'(0)=-{3\over2}\ze_R'(-2)+{1\over6}\ln2+{11\over48},\cr
\noalign{\vskip5truept}
&\ze_4'(0)={2\over3}\big(\ze_R'(-3)-\ze_R'(-1)\big)+
{1\over45}\ln2-{2489\over30240},\cr
\noalign{\vskip5truept}
&\ze_5'(0)=
%{3\over16}\ze_R'(0,1/2)-{5\over6}\ze_R'(-2,1/2)
%+{1\over3}\ze_R'(-4,1/2)+{17\over1440}\ln2-{17497\over241920}
\frac 5 8 \zeta_R'(-2)-\frac 5 {16} \zeta_R' (-4) -\frac{59}{720}\ln 2
-\frac{17497}{241920},\cr
\noalign{\vskip5truept}
&\ze_6'(0)={4\over15}\ze_R'(-1)-{1\over3}\ze_R'(-3)+{1\over15}\ze_R'(-5)
-{1\over189}\ln2+{6466519\over207567360},\cr
\noalign{\vskip5truept}
&\ze_7'(0)=
%-{5\over64}\ze_R'(0,1/2)+{259\over720}\ze_R'(-2,1/2)
%-{7\over36}\ze_R'(-4,1/2)\cr
%&\hspace{****************}+{1\over45}\ze_R'(-6,1/2)-{367\over120960}\ln2
%-{59792179\over2075673600}
-\frac{259}{960}\zeta_R'(-2) +\frac{35}{192}\zeta_R'(-4)-\frac 7 {320}
\zeta_R'(-6) +\frac{2179}{60480}\ln 2
+\frac{59792179}{2075673600},\cr
\noalign{\vskip5truept}
&\ze_8'(0)=-{4\over35}\ze_R'(-1)+{7\over45}\ze_R'(-3)-{2\over45}\ze_R'
(-5)\cr
&\hspace{****************}+{1\over315}\ze_R'(-7)+{23\over14175}\ln2
-{183927381289\over14079294028800}.\cr}
\eql{sdet}$$
The four dimensional result is that already computed in
[\pref{KandC,Dow10}].

\noin{\bf6. Mixed functional determinants.}

The mixed determinants are likewise found to be given in terms of
$$\eqalign{
&\ze_2'(0)=2\ze_R'(-1)+{1\over6}\ln2-{1\over12},\cr
\noalign{\vskip5truept}
&\ze_3'(0)=-{3\over2}\ze_R'(-2)+{1\over4}\ln2+{1\over16},\cr
\noalign{\vskip5truept}
&\ze_4'(0)={251\over15120}-{11\over180}\ln2
+{2\over3}\big(\ze_R'(-3)-\ze_R'(-1)\big),\cr
\noalign{\vskip5truept}
&\ze_5'(0)=-{91\over3840}-{3\over32}\ln 2 -{5\over16}\ze_R'(-4)
+{5\over8}\ze_R '(-2),\cr
\noalign{\vskip5truept}
&\ze_6'(0)=-{28417\over4989600} +{191\over7560}\ln 2
+{1\over15}\ze_R'(-5) -{1\over3}\ze_R'(-3) +{4\over15}\ze_R' (-1),\cr
\noalign{\vskip5truept}
&\ze_7'(0)={47941\over4838400}+{5\over128}\ln 2 -{7\over320}
\ze_R'(-6) +{35\over192}\ze_R'(-4)-{259\over960}\ze_R'(-2),\cr
\noalign{\vskip5truept}
&\ze_8'(0)={14493407\over6399679104}-{2497\over226800}\,\ln2+
{1\over315}\ze_R'(-7)-{2\over45}\,\ze_R'(-5)\cr
\noalign{\vskip5truept}
&\hspace{*************************}+{7\over45}\,\ze_R'(-3)
-{4\over35}\,\ze_R'(-1).\cr}
\eql{ldet}$$

It is worth noting that the two- three- and four-dimensional results
agree with those found
by one of us (JSA) using a conformal transformation method [\pref{DandA2}].
Again, the four dimensional result is that given in [\pref{KandC,Dow10}].
\section{\bf7. Conclusion.}
As noted earlier, the specific expressions obtained here may be of use
in tying down the general form of the heat-kernel coefficients in the
spectral case, if there is one. Grubb and Seeley's [\pref{GandS1}] formal
results on the expansion have already been alluded to. The
work of Gilkey [\pref{Gilkey2}] is mostly concerned with that combination of
coefficients relevant for the spin index.

More might be said for the mixed coefficients. For example, the general
form of the mixed $B_{5/2}$ could be written down following
[\pref{BandG2}]. Then, specialising to a flat ambient manifold,
precise values for some coefficients, and for combinations of others,
could be obtained in the manner of van den Berg (reported in
[\pref{BGV}]) who used the Dirichlet scalar polynomials computed by Levitin
[\pref{Levitin}].
This programme will be pursued elsewhere. Unfortunately the procedure will
not be as informative as the corresponding scalar one, where one
has the extra control provided by the Robin multiplier. For example, using the
polynomials derived by Levitin, the following Neumann coefficients in lemma
5.1 of [\pref{BGV}] are very easily obtained,
$$\eqalign{
&d_{30}=2160,\quad d_{31}=1080,\quad d_{32}=360,\quad d_{33}={885\over4},
\quad d_{34}={315\over2}\cr
\noalign{\vskip7truept}
&d_{35}=150,\,d_{36}={2041\over128},\,d_{37}={417\over32},\,
d_{38}+d_{39}={1175\over32},\,d_{40}={231\over8}.\cr}
$$
\section{\bf8. Acknowledgments.}

The work of KK is supported by the DFG under contract number Bo 1112/4-1.
JSA would like to thank the ESPRC for a research studentship.
\newpage
\section{\bf Appendix A. $\bga$--matrices and spinors.}
In a $d$--dimensional space, we denote by $\ga^a_{(d)}$ , $a=1,2,\ldots d$,
the $\ga$--matrices projected along some $d$--bein system.
If $d$ is even, the $\ga$'s are defined inductively by
$$\eqalign{
&\ga^j_{(d)}=\left(\matrix{{\bf0}&i\ga^j_{(d-2)}
\cr-i\ga^j_{(d-2)}&{\bf 0}}\right),\quad j=1,2,\ldots d-1,\cr
\noalign{\vskip10truept}
&\ga^d_{(d)}=\left(\matrix{{\bf 0}&{\bf1}\cr{\bf1}&{\bf0}}\right),\qquad
\ga^{d+1}_{(d)}=\left(\matrix{{\bf1}&0\cr{\bf0}&-{\bf1}}\right)\cr}
\eql{egammas}$$starting from the Pauli matrices
$$
\ga^1_{(2)}=\left(\matrix{0&i\cr-i&0}\right),\quad
\ga^2_{(2)}=\left(\matrix{0&1\cr1&0}\right),\quad
\ga^3_{(2)}=\left(\matrix{1&0\cr0&-1}\right).
$$
The matrices (\peq{egammas}) satisfy the Dirac anti-commutation formula
$$
\ga^a_{(d)}\,\ga^b_{(d)}+\ga^b_{(d)}\,\ga^a_{(d)}=2\de^{ab}.
$$

In the body of this paper on the $d$-ball, $\ga^{d+1}_{(d)}$ is denoted by
$\Ga^5$ and $\ga^d_{(d)}$ by $\Ga^r$, the (outward) radial matrix. For
example, the mixed projector (\peq{mproj}) is written here as
$$
P_+\psi={1\over2}\big({\bf1}-i\ga^{d+1}_{(d)}\ga^a_{(d)}e^\mu_a
\,n_\mu\big)\psi=0
$$ where $e^\mu_a$ is the $d$--bein.

For spectral conditions, in the terminology of [\pref{Gilkey2}],
$\widetilde\ga^j_{(d-1)}=i\ga^d_{(d)}\,\ga^j_{(d)}$ is the {\it induced
tangential} Clifford module structure on the boundary confined spinor
bundle and satisfies
$$
\widetilde\ga^i_{(d-1)}\,\widetilde\ga^j_{(d-1)}
+\widetilde\ga^j_{(d-1)}\,\widetilde\ga^i_{(d-1)}=2\de^{ij}.
$$

In the present work, the matrices for odd $d$ are defined in terms of
those for even $d$ in the following way,
$$\eqalign{
&\ga^j_{(d)}=\ga^j_{(d+1)},\quad j=1,2,\ldots d-1,\cr
\noalign{\vskip10truept}
&\ga^d_{(d)}=\ga^{d+1}_{(d+1)},\qquad\ga^{d+1}_{(d)}=\ga^{d+2}_{(d+1)},\cr}
\eql{ogamma}$$
and $\ga^d_{(d+1)}$ is not used.
Again, $\ga^d_{(d)}$ is the radial matrix and $\ga^{d+1}_{(d)}$ is
`$\,\Ga^5\,$'.
This particular choice has the advantage of giving the same mode structure
in both odd and even dimensions.

In effect, we are defining spinors on odd $\man$ through those on even
$\oR\times\man$ by ignoring the added dimension, \eg by taking fields
uniform on the $\oR$.

Of course this is what physicists have done automatically from the first
when separating variables for the Dirac equation in, say, polar
coordinates. A pertinant case is the Casimir energy in a spatial 3-sphere
in Minkowski space-time \eg [\pref{Milton, BandH, BandI}].

The use of doubled $\ga$-matrices for odd dimensions in the present paper was
motivated originally by the desire to implement mixed boundary conditions,
for which $d$ matrices are needed to contract into the normal plus one
further matrix that anti-commutes with these. Since there are not enough
matrices in the usual irreducible representation (of dimension $2^{(d-1)/2}$)
of the Clifford-Dirac algebra to accomplish this, the dimension was
doubled and $\ga$-matrices of one {\it higher} dimension used, with a
single redundancy. This means, for example, that 2-spinors can be defined
on the 3-sphere, but not on the 3-hemisphere.
These doubled-up matrices also allow one to discuss spectral
conditions for odd $d$, as in the text. Another approach, using
pin manifolds, is discussed by Gilkey, [\pref{Gilkey2}] section 9.

Trautman [\pref{Trautman}] refers to spinors in
$\oR^{2n}$ as {\it Dirac} spinors, which, when restricted to a hypersurface,
become {\it Cartan} spinors. There seems to be no reason why this
terminology cannot be extended to curved spaces.
\newpage
\section{\bf Appendix B. Mixed coefficient polynomials.}

The mixed coefficients have the structure,
$$\eqalign{
B_n^{(L)}(d)&={2^{-d}d_s\over\Ga(d/2)}\sqrt\pi\,(d-1)\,P_n(d),
\quad n=1/2,3/2,\ldots\cr
\noalign{\vskip4truept}
&={2^{-d}d_s\over\Ga(d/2)}\,(d-1)\,P_n(d),\quad n=1,2,\ldots,\cr}
\eql{lform2}$$
with the polynomials
$$\eqalign{
&P_{5/2}^{(L)}(d)={1\over122880}
\,\left(d+1\right)\,\left(d-5\right)\,\left(89\,d-263\right),\cr
\noalign{\vskip5truept}
&P_3^{(L)}(d)=-{1\over1247400}\,
\left(15600+11426\,d-9169\,{d^2}+1006\,{d^3}+61\,{d^4}\right),\cr
\noalign{\vskip5truept}
&P_{7/2}^{(L)}(d)={1\over495452160}\,\left(d-7\right)\cr
\noalign{\vskip4truept}
&\hspace{*******}\left(393039+368952\,d-147742\,{d^2}-33848\,
{d^3}+9167\,{d^4}\right),\cr
\noalign{\vskip5truept}
&P_4^{(L)}(d)=-{1\over219988969200}\,\left(1908965520+1529812932\,
d-808656824\,{d^2}\right.\cr
\noalign{\vskip4truept}
&\hspace{*********}\left.-197908917\,{d^3}+105046309\,{d^4}-10068831\,
{d^5}+83899\,{d^6}\right),\cr
\noalign{\vskip5truept}
&P_{9/2}^{(L)}(d)={1\over20927899238400}
\,\left(d+1\right)\,\left(d-9\right)\cr
\noalign{\vskip4truept}
&\hspace{*******}\left(10887720195-916876245\,d-2084061206\,{d^2}
+333544346\,{d^3}\right.\cr
\noalign{\vskip4truept}
&\hspace{**********************}\left.+40853459\,{d^4}-6852869\,{d^5}
\right),\cr
\noalign{\vskip5truept}
&P_5^{(L)}(d)=-{1\over6830657493660000}\cr
\noalign{\vskip5truept}
&\hspace{*******}\left(57920260204800+47074221218160\,d-20614444675524
\,{d^2}\right.\cr
\noalign{\vskip4truept}
&\hspace{*********}-7939793557052\,{d^3}+2539767459817\,{d^4}
+254749941880\,{d^5}\cr
\noalign{\vskip4truept}
&\hspace{**************}\left.-118154075186\,{d^6}+9525728692\,{d^7}
-170628227\,{d^8}\right).\cr}
$$
\newpage
\noin{\bf Appendix C. Spectral coefficient polynomials.}
$$\eqalign{
&F_{5/2}(d)=-{1\over {60480}}\,({46809-27899\,d+4536\,{d^2}-160\,{d^3}}),\cr
&G_{5/2}(d)={1\over{368640}}\,({9927-5129\,d+369\,{d^2}+65\,{d^3}}),\cr
&F_3(d)=-{1\over{405405}}\,{\left(d-8\right)\,
\left(1542-385\,d-171\,{d^2}+40\,{d^3}\right)},\cr
&G_3(d)={1\over{737280}}\,{\left(d-8\right)\,
\left(63600 -33668\,d+3924\,{d^2}+65\,{d^3}\right)},\cr
&F_{7/2}(d)=-{1\over{103783680}}\,\big(221818311-156858900\,d+35468617\,
{d^2}\cr
&\hspace{**************}-2592500\,{d^3}-24928\,{d^4}+5120\,{d^5}\big),\cr
\noalign{\vskip3truept}
&G_{7/2}(d)={1\over{371589120}}\,(4501359-827409\,d-1050058\,{d^2}
+372374\,{d^3}\cr
&\hspace{***************************}-35141\,{d^4}+475\,{d^5}),\cr
&F_4(d)=-\frac1{13749310575}\left(d-10\right)\,
\left(23041368 + 2531082\,d\right.\cr
&\hspace{*************}\left.- 8288995\,{d^2} + 1680941\,{d^3}
+24355\,{d^4}-16775\,{d^5}\right),\cr
&G_4(d)=\frac 1 {743178240}
\left( d -10\right) \,\left( 170021376 - 111709248\,d\right.\cr
\noalign{\vskip3truept}
& \hspace{*************}\left.+21793760\,{d^2}-1009228\,{d^3}-50096\,{d^4}
+475\,{d^5}\right),\cr
&F_{9/2}(d)=-\frac1{56317176115200}\,(464260690378485-373244849131275\,d\cr
\noalign{\vskip3truept}
&\hspace{********}+ 106164603742547\,{d^2} - 12690585476317\,
{d^3}+504841197392\,{d^4}\cr
&\hspace{*************}+7017579968\,{d^5}-26306560
\,{d^6}-34355200\,{d^7}),\cr
&G_{9/2}(d)=\frac 1 {62783697715200}\,(509445573615 + 91281582927\,d\cr
\noalign{\vskip3truept}
&\hspace{*************}- 236987179165\,{d^2} + 54541934915\,{d^3}
+2269235885\,{d^4}\cr
&\hspace{***************}-1736240947\,{d^5}+155559425\,{d^6}-3457375\,
{d^7}),\cr
&F_5(d)=-\frac1{8538321867075}\,\big(d-12\big)\,\big(9567536832+3811378020
\,d\cr
\noalign{\vskip3truept}
&\hspace{*************}-4614231340\,{d^2}+405754883\,{d^3}+230777003\,
{d^4}\cr
&\hspace{***************}- 42837163\,{d^5}+1236545\,{d^6}+84980\,
{d^7}\big),\cr
&G_5(d)=\frac1{125567395430400}\left(d-12\right)\,
\left( 105020227952640 - 80282869575168\,d \right.\cr
\noalign{\vskip3truept}
&\hspace{***********}+20966354815040\,{d^2}-2131292479600\,{d^3}
+52372511840\,{d^4}\cr
&\hspace{**************}+1363355768\,{d^5}\left.+125123300\,{d^6}-3457375\,
{d^7}\right).\cr}
$$
\newpage

\section{\bf References}
\vskip 5truept
\begin{putreferences}
\ref{BGKE}{M.Bordag, B.Geyer, K.Kirsten and E.Elizalde, {\it Zeta function
determinant of the Laplace operator on the D-dimensional ball},
hep-th/9505157, {\it Comm. Math. Phys.} to be published.}
\ref{KandC}{K.Kirsten and G.Cognola, {\it Heat-kernel coefficients and
functional determinants for higher spin fields on the ball} UTF354. Aug.
1995, hep-th/9508088.}
\ref{Dow8}{J.S.Dowker {\it Robin conditions on the Euclidean ball}
MUTP/95/7; hep-th\break/9506042.}
\ref{Dow9}{J.S.Dowker {\it Oddball determinants} MUTP/95/12; hep-th/9507096.}
\ref{Dow10}{J.S.Dowker {\it Spin on the 4-ball},
hep-th/9508082, {\it Phys. Lett. B}, to be published.}
\ref{DandA2}{J.S.Dowker and J.S.Apps, {\it Functional determinants on
certain domains}. To appear in the Proceedings of the 6th Moscow Quantum
Gravity Seminar held in Moscow, June 1995; hep-th/9506204.}
\ref{BEK}{M.Bordag, E.Elizalde and K.Kirsten {\it Heat kernel
coefficients of the Laplace operator on the D-dimensional ball},
hep-th/9503023, {\it J.Math.Phys.}, to be published.}
\ref{DandH}{P.D.D'Eath and J.J.Halliwell \prD{35}{87}{1100}.}
\ref{DandE}{P.D.D'Eath and G.V.M.Esposito \prD{43}{91}{3234}.}
\ref{DandE2}{P.D.D'Eath and G.V.M.Esposito \prD{44}{91}{1713}.}
\ref{Moss}{I.G.Moss \cqg{6}{89}{659}.}
\ref{MandP}{I.G.Moss and S.J.Poletti \pl{B333}{94}{326}.}
\ref{Barv}{A.O.Barvinsky, Yu.A.Kamenshchik and I.P.Karmazin \aop {219}
{92}{201}.}
\ref{Kam}{Yu.A.Kamenshchik and I.V.Mishakov \prD{47}{93}{1380}.}
\ref{KandM}{Yu.A.Kamenshchik and I.V.Mishakov {\it Int. J. Mod. Phys.}
{\bf A7} (1992) 3265.}
\ref{MandP2}{I.G.Moss and S.J.Poletti \np{341}{90}{155}.}
\ref{Luck}{H.C.Luckock \jmp{32}{91}{1755}.}
\ref{Poletti}{S.J.Poletti \pl{B249}{90}{355}.}
\ref{Vass}{D.V.Vassilevich.{\it Vector fields on a disk with mixed
boundary conditions} gr-qc /9404052.}
\ref{Kam2}{G.Esposito, A.Y.Kamenshchik, I.V.Mishakov and G.Pollifrone
\prD{50}{94}{6329}; \prD{52}{95}{3457}.}
\ref{MandV}{V.N.Marachevsky and D.V.Vassilevich {\it Diffeomorphism
invariant eigenvalue \break problem for metric perturbations in a bounded
region}, SPbU-IP-95, \break gr-qc/9509051.}
\ref{Esposito}{G.Esposito {\it Quantum Gravity, Quantum Cosmology and
Lorentzian Geometries}, Lecture Notes in Physics, Monographs, Vol. m12,
Springer-Verlag, Berlin 1994.}
\ref{Esposito2}{G.Esposito {\it Nonlocal properties in Euclidean Quantum
Gravity}. To appear in Proceedings of 3rd. Workshop on Quantum Field Theory
under the Influence of External Conditions, Leipzig, September 1995;
gr-qc/9508056.}
\ref{BGV}{T.P.Branson, P.B.Gilkey and D.V.Vassilevich {\it The Asymptotics
of the Laplacian on a manifold with boundary} II, hep-th/9504029.}
\ref{APS}{M.F.Atiyah, V.K.Patodi and I.M.Singer \mpcps{77}{75}{43}.}
\ref{FORW}{P.Forgacs, L.O'Raifeartaigh and A.Wipf \np{293}{87}{559}.}
\ref{Gilkey2}{P.B.Gilkey \aim{102}{93}{129}.}
\ref{Gilkey3}{P.B.Gilkey \aim{15}{75}{334}.}
\ref{NandT}{M.Ninomiya and C.I.Tan \np{245}{85}{199}.}
\ref{Ma}{Z.Q.Ma \jpa{19}{86}{L317}.}
\ref{MandS}{A.V.Mishchenko and Yu.A.Sitenko \aop{218}{92}{199}.}
\ref{RandS}{H.R\"omer and P.B.Schroer \pl{21}{77}{182}.}
\ref{Trautman}{A.Trautman \jmp{33}{92}{4011}.}
\ref{Milton}{K.A.Milton \prD{22}{80}{144}.}
\ref{BandH}{C.M.Bender and P.Hays \prD{14}{76}{2622}.}
\ref{BandI}{J.Baacke and Y.Igarishi \prD{27}{83}{460}.}
\ref{GandS1}{G.Grubb and R.T.Seeley \cras{317}{1993}{1124}; \invm{121}{95}
{481}.}
\ref{NandS}{A.J.Niemi and G.W.Semenoff \np {269}{86}{131}.}
\ref{ETP}{G.Esposito, H.A.Morales-T\'ecotl and L.O.Pimentel {\it Essential
self-adjointness in one-loop quantum cosmology}, gr-qc/9510020.}
\ref{MandL}{H.C.Luckock and I.G.Moss \cqg{6}{89}{1993}.}
\ref{BandG2}{T.P.Branson and P.B.Gilkey {\it Comm. Partial Diff. Eqns.}
{\bf 15} (1990) 245.}
\ref{Grubb}{G.Grubb {\it Comm. Partial Diff. Eqns.} {\bf 17} (1992)
2031.}
\ref{AandT}{M.A.Awada and D.J.Toms \np{245}{84}{161}.}
\ref{JandK}{T.Jaroszewicz and P.S.Kurzepa \aop{213}{92}{135}.}
\ref{CandH}{R.Camporesi and A.Higuchi {\it On the eigenfunctions of the
Dirac operator on spheres and real hyperbolic spaces}, gr-qc/9505009.}
\ref{Levitin}{M.Levitin {\it Dirichlet and Neumann invariants for Euclidean
balls}, {\it Diff. Geom. and its Appl.}, to be published.}
\ref{Mcav}{D.M.Mcavity \cqg{9}{92}{1983}.}
\end{putreferences}
\bye